\documentclass[letterpaper, 12pt]{article}
\usepackage{graphicx,amssymb}
\usepackage[usenames]{color}
\usepackage[small,compact,nonindentfirst]{titlesec}
\usepackage[square,numbers,sort&compress]{natbib}
\usepackage{natbib,multicol}
\usepackage{wrapfig}

\newcommand{\remove}[1]{}

\newcommand{\altaffilmark}[1]{$^{#1}$}
\newcommand{\altaffiltext}[2]{{\noindent \small $^{#1}$ #2} \\}

\titlespacing{\section}{0pt}{*1}{0.5ex plus 0.2ex minus 0.2ex}
\titlespacing{\subsection}{0pt}{0.7ex plus 0.2ex minus 0.2ex}
{0.5ex plus 0.2ex minus 0.2ex}
\newcommand{\captionfonts}{\small}

\makeatletter  
\long\def\@makecaption#1#2{%
  \vskip\abovecaptionskip
  \sbox\@tempboxa{{\captionfonts #1: #2}}%
  \ifdim \wd\@tempboxa >\hsize
    {\captionfonts #1: #2\par}
  \else
    \hbox to\hsize{\hfil\box\@tempboxa\hfil}%
  \fi
  \vskip\belowcaptionskip}
\makeatother   

\long\def\symbolfootnote[#1]#2{\begingroup%
\def\thefootnote{\fnsymbol{footnote}}\footnote[#1]{#2}\endgroup}

\begin{document}
\pagestyle{empty}

\begin{center}
{\Large \bf A New Era in Extragalactic Background Light Measurements:  \\[0.8ex] 
\large The Cosmic History of Accretion, Nucleosynthesis and Reionization}
\end{center}

\vspace{1.0ex}
{Asantha Cooray\altaffilmark{1,}\symbolfootnote[1]{{\it E-mail: acooray@uci.edu; Tel: 949-701-6393}}
Alexandre Amblard\altaffilmark{1},
Charles Beichman\altaffilmark{2},
Dominic Benford\altaffilmark{3}, 
Rebecca Bernstein\altaffilmark{4},
James~J. Bock\altaffilmark{2,5},
Mark Brodwin\altaffilmark{6},
Volker Bromm\altaffilmark{7},
Renyue Cen\altaffilmark{8},
Ranga R. Chary\altaffilmark{2},
Mark Devlin\altaffilmark{9},
Timothy Dolch\altaffilmark{10},
Herv\'e Dole\altaffilmark{11},
Eli Dwek\altaffilmark{3},
David Elbaz\altaffilmark{12},
Michael Fall\altaffilmark{10},
Giovanni Fazio\altaffilmark{13},
Henry Ferguson\altaffilmark{10},
Steven Furlanetto\altaffilmark{14},
Jonathan Gardner\altaffilmark{3},
Mauro Giavalisco\altaffilmark{15},
Rudy Gilmore\altaffilmark{4},
Nickolay Gnedin\altaffilmark{16},
Anthony Gonzalez\altaffilmark{17},
Zoltan Haiman\altaffilmark{18},
Michael Hauser\altaffilmark{9},
Jiasheng Huang\altaffilmark{13},
Sergei Ipatov\altaffilmark{19},
Alexander Kashlinsky\altaffilmark{3},
Brian Keating\altaffilmark{20}, 
Thomas Kelsall\altaffilmark{3},
Eiichiro Komatsu\altaffilmark{7},
Guilaine Lagache\altaffilmark{11},
Louis R. Levenson\altaffilmark{2},
Avi Loeb\altaffilmark{13},
Piero Madau\altaffilmark{4},
John C. Mather\altaffilmark{3},
Toshio Matsumoto\altaffilmark{21},
Shuji Matsuura\altaffilmark{21},
Kalevi Mattila\altaffilmark{22},
Harvey Moseley\altaffilmark{3},
Leonidas Moustakas\altaffilmark{5},
S. Peng Oh\altaffilmark{23},
Larry Petro\altaffilmark{10},
Joel Primack\altaffilmark{4},
William Reach\altaffilmark{2},
Tom Renbarger\altaffilmark{20},
Paul Shapiro\altaffilmark{7},
Daniel Stern\altaffilmark{5}, 
Ian Sullivan\altaffilmark{2}, 
Aparna Venkatesan\altaffilmark{24},
Michael Werner\altaffilmark{5},
Rogier Windhorst\altaffilmark{25},
Edward L. Wright\altaffilmark{14},
Michael Zemcov\altaffilmark{2,5}
}

\vspace{0.15in}
\altaffiltext{1}{Center for Cosmology, University of California, Irvine, CA 92697}
\altaffiltext{2}{IPAC/Physics/Astronomy, California Institute of Technology, Pasadena, CA 91125}
\altaffiltext{3}{NASA/GSFC, Code 665, Greenbelt, MD 20771}
\altaffiltext{4}{Department of Astronomy \& Astrophysics, University of California, Santa Cruz, CA 95064}
\altaffiltext{5}{Jet Propulsion Laboratory, 4800 Oak Grove Drive, Pasadena, CA 91109}
\altaffiltext{6}{NOAO, Tucson, AZ 85719}
\altaffiltext{7}{Department of Astronomy, University of Texas, Austin, TX 78712}
\altaffiltext{8}{Department of Astrophysical Sciences, Princeton University, Princeton, NJ 08544}
\altaffiltext{9}{Department of Physics, University of Pennsylvania, Philadelphia, PA 19104}
\altaffiltext{10}{STScI, 3700 San Martin Dr., Baltimore, MD 21218}
\altaffiltext{11}{IAS, Universit\'e Paris, Orsay Cedex, France}
\altaffiltext{12}{CEA Saclay, Service d'Astrophysique, Gif-sur-Yvette Cedex, France}
\altaffiltext{13}{Center for Astrophysics, 60 Garden Street, Cambridge, MA 02138}
\altaffiltext{14}{UCLA Physics \& Astronomy, Los Angeles, CA 90095}
\altaffiltext{15}{Department of Astronomy, University of Massachusetts, Amherst, MA 01003}
\altaffiltext{16}{Theoretical Astrophysics Group, Fermilab, Batavia, IL 60510}
\altaffiltext{17}{Department of Astronomy, University of Florida, Gainesville, FL 32611}
\altaffiltext{18}{Department of Astronomy, Columbia University, New York, NY 10027}
\altaffiltext{19}{Department of Physics, Catholic University of America, Washington, DC 20064}
\altaffiltext{20}{Department of Physics, University of California, La Jolla, CA 92093}
\altaffiltext{21}{ISAS, JAXA, Sagamihara, Kanagawa 229-8510, Japan} 
\altaffiltext{22}{Observatory, University of Helsinki, Helsinki, Finland}
\altaffiltext{23}{Department of Physics, University of California, Santa Barbara, CA 93106}
\altaffiltext{24}{Department of Physics \& Astronomy, University of San Francisco, San Francisco, CA 94117} 
\altaffiltext{25}{Department of Physics, Arizona State University, Tempe AZ 85287}

\vspace{0.5in}

\newpage
\pagestyle{plain}
\setcounter{page}{1}

\begin{center}
{\bf Executive Summary}
\end{center}

{\bf What is the total radiative content of the Universe since the epoch of recombination?} 
The extragalactic background light (EBL) spectrum captures the redshifted energy released from the first stellar objects, 
protogalaxies, and galaxies throughout cosmic history. 
It is a key constraint on all models of galaxy formation and evolution, and provides an anchor that connects
global radiation energy density to star formation, metal production, and gas consumption. Yet, we have not 
determined the brightness of the extragalactic sky
from UV/optical to far-infrared wavelengths with sufficient accuracy to establish the radiative content of the Universe
to better than an order of magnitude. It was due to the first-generation of EBL measurements that we now know
that the far-infrared background associated with dust in high redshift galaxies is energetically as important as the
optical/near-IR background. This was an unexpected discovery.

As a function of wavelength, a cosmic 
consistency test can be performed by comparing the integrated light from all galaxies resolved 
by both ground and space-based observatories 
to the EBL intensity. Any discrepancies suggest the presence of new, diffuse sources unresolved by telescopes.
The possibilities for new discoveries with profound implications for astronomy range from recombination signatures during 
reionization, unexpected sources such as primordial black holes,
photons from decay of elementary particles, to a new component of the interstellar medium of the Milky Way.

Among many science topics, an accurate measurement of the EBL spectrum from optical to far-IR wavelengths, will address
the following questions:
\begin{itemize}
\item What is the total energy released by stellar nucleosynthesis over cosmic history?
\item Was significant energy released by non-stellar processes?
\item Is there a diffuse component to the EBL anywhere from optical to sub-millimeter?
\item When did first stars appear and how luminous was the reionization epoch?
\end{itemize}

Zodiacal  dust in the Solar System is the main foreground that limits EBL measurements at optical/near-IR 
wavelengths, while interstellar dust limits measurements at sub-mm wavelengths. New EBL measurements must be performed
in parallel with observations that further our understanding of zodiacal and Galactic dust. 
Absolute optical to mid-IR EBL spectrum to an astrophysically interesting accuracy can be established
by wide field imagingat a distance of 5 AU or above the ecliptic plane where the zodiacal 
foreground is reduced by more than two orders of magnitude.
Such an imaging opportunity could be conceived as part of a planetary mission to the outer Solar System.

The high-energy source community will benefit from a direct measurement of the EBL spectrum as it allows the intrinsic
spectra of AGNs at TeV energies to be established. This will enable studies on
 time variability and the acceleration of relativistic electrons responsible for the TeV emission. 
The planetary scientists  and dynamicists will benefit from detailed measurements of the zodiacal dust cloud as details of its origin  and the source of 
zodiacal dust are still hotly debated. A detailed characterization of our zodiacal cloud will provide critical information
required to fine-tune imaging studies of extra-Solar terrestrial planets and exo-zodiacal clouds around nearby stars.

\newpage

{\it Introduction:} 
A complete understanding of the total energy content of the Universe across the entire
electromagnetic spectrum is still lacking. 
While at radio wavelengths the cosmic microwave background (CMB) produced by primordial photons has been well studied over 
the last quarter century, 
with a detailed characterization of CMB fluctuations with two satellite missions (COBE and WMAP),
the same cannot be said of the backgrounds that peak at both optical/near-IR and far-IR (submm) wavelengths.
At optical and near-IR wavelengths the absolute EBL is a measure of the energetics of unobscured star formation in the Universe [1], 
while at far-IR wavelengths the EBL gives a measure of the photon content reprocessed by dust, both in galaxies and 
between galaxies [2]. Thus, the near-IR EBL has been suggested as a way to identify first sources that reionized the Universe while the 
sub-mm EBL reveals the star-formation history enshrouded by dust.

{\bf What is the total energy released by nucleosynthesis over cosmic history?} Our attempts to resolve the EBL into individual galaxies rely on deep observations which measure
the contributions of faint sources. At far-infrared wavelengths where source confusion due to the large point
spread function is an issue, we predominantly rely on stacking homogeneous populations of galaxies detected at other
passbands to measure the contribution of faint, hitherto undetected sources. 
Both these techniques suggest that $\sim$85\% of the integrated galaxy light arises from galaxies at $z<1.5$ corresponding to 
the peak in the star-formation history and the rapid growth of stellar mass in galaxies [3].

However, integrated galaxy light estimates typically fall short of the absolute EBL
values. For instance at 3.6 $\mu$m, the galaxies contribute an intensity of $\sim6$-9 nW~m$^{-2}$~sr$^{-1}$ [4]. In
contrast, the EBL measured by DIRBE at the same wavelength is (13.3$\pm$2.8) nW~m$^{-2}$~sr$^{-1}$ [5]. 
Similarly, at FIR wavelengths, the stacking at 160$\mu$m reveals a galaxy contribution  of (13.4$\pm$1.7) nW~m$^{-2}$~sr$^{-1}$ [6] while the
DIRBE measured EBL is (25$\pm$7) nW~m$^{-2}$~sr$^{-1}$ [7]. It is not yet clear if there is a
EBL contribution  from star-formation and AGN activity which are hidden from traditional deep, pencil-beam surveys, 
particularly at $z>1.5$. Such a revelation would dramatically change our understanding of the peak of cosmic star-formation
and indicate the presence of massive, quiescent (or dust obscured) galaxies at $z>1.5$ which are missed in traditional, Lyman-break galaxy studies.

Furthermore, a discovery of hidden star-formation and stellar mass,
would help address the difficulties associated with keeping the intergalactic medium in the Universe ionized
between z$\sim3-6$. Current models require a non-Salpeter, top-heavy IMF [8] in galaxies
at $z>2$ to reconcile estimates of star-formation rates and stellar mass density 
estimates and for maintaining the ionized state of the IGM.

The EBL spectrum also contains all radiative information from the reionization epoch [9]. Diffuse signatures are expected in terms of
 Ly-$\alpha$  background radiation redshifted to near-IR wavelengths today.
While individual galaxies contributing to such a background during reionization will be most likely resolved 
by JWST, due to their large clumping factors, any backgrounds that scatter 
to a larger extent especially in an under dense region will contribute to the EBL [10], but most likely not be detected by JWST. 
Whether the epoch of dust formation quickly followed reionization or whether that epoch was substantially delayed since
the appearance of first stars holds clues to galaxy formation and first-light sources. A careful analysis of the sub-millimeter background 
combined with galaxy counts and fluctuations of the EBL at near-IR from JWST and future IR missions can address this question.

\begin{figure*}[!t]
\begin{center}
\includegraphics[width=6.5in]{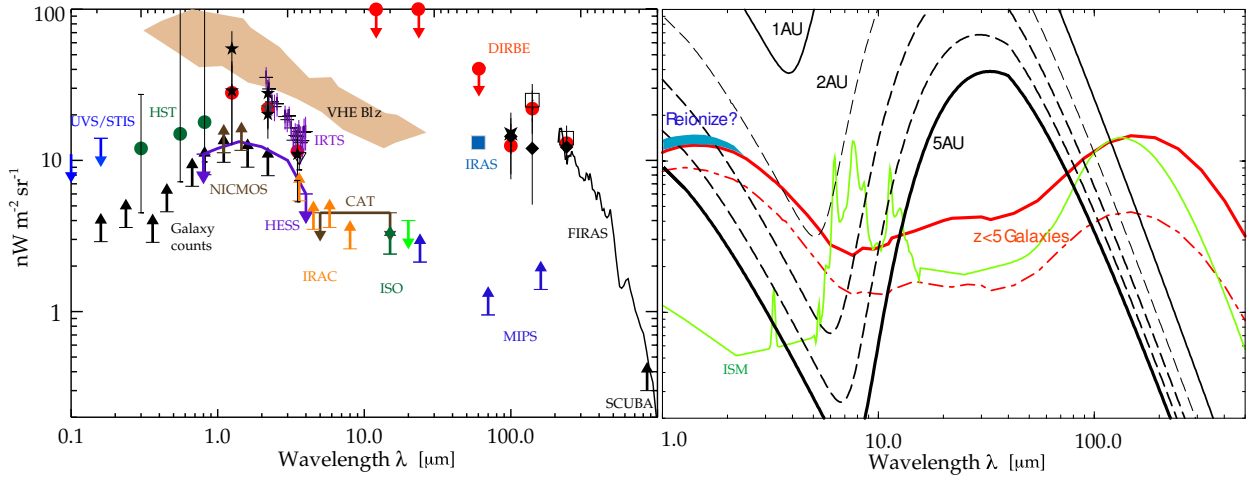}
\end{center}
\caption{
{\it Left panel:} The total optical to far-IR EBL measured from a combination of absolute photometry (ie. HST, DIRBE, IRTS, and FIRAS) and with indirect techniques  
(GeV/TeV blazar spectra and resolved/stacked galaxy counts) [6]. 
{\it Right panel:} The scattered light and thermal emission (black lines) at 1 AU to 5 AU from the Sun, for the
zodiacal cloud with a radial density profile of $r^{-3/2}$. The line labeled ISM is the Galaxy emission
towards a typical low-dust field. The curves in red show two predictions for the integrated light from $z < 5$ galaxies in semi-analytical models 
normalized to observed galaxy counts (with pre-2005 semi-analytic models in dot-dashed). The shaded region labeled ``Reionize?'' shows the wavelength range of interest for EBL
 measurements targetting diffuse emission from reionization, such as redshifted Lyman-$\alpha$.
 At a distance of 5 AU, absolute photometry measurements can be carried out to determine the EBL spectral signature from reionization, 
without any significant zodiacal foreground as at 1 AU.
}
\label{fig:ebl}
\end{figure*}

{\bf Was significant energy released by non-stellar processes?}
The dust torii surrounding active galactic nuclei leads to a secondary contribution from
AGN to the EBL at longer wavelengths, providing a way to study the role played
by AGN in  galaxy formation and evolution. Contributions to the EBL spectrum are also expected from
 first stars and primordial blackholes in the form of miniquasars during reionization [11] and
diffuse emission from haloes of galaxies and clusters. The reionization of the diffuse
intergalactic medium will be accompanied by radiative recombination radiation with Lyman-$\alpha$ line from collisional
excitation. An exciting possibility is a diffuse background from elementary particle decays. Neutralinos that decay to
 gravitinos and photons with mass splittings in the order of eV, instead of the usual MeV products,
could lead to a photon background at IR wavelengths [12]. The lifetimes would be greater than 10 Gyrs 
and the photons would cluster similar to dark matter at late times instead of tracing luminous sources at halo centers.

In the future, sub-millimeter wavelength absolute and fluctuation measurements can also be used to constrain any spectral 
distortions to the CMB associated with most mechanisms of
energy generation during reionization.  In the wavelength range 100 $\mu$m to 1000 $\mu$m broad
recombination lines from cosmological H and He  before dark ages  are also expected
with intensities at the level of $10^{-5}$ nW m$^{-2}$ sr$^{-1}$ [13].

{\it State of the field:} In Figure~1 (left panel) we summarize EBL intensity measurements between optical and far-IR wavelengths using
absolute photometry, the integrated galaxy light from galaxy counts (lower limits), or with  the TeV spectra of high redshift sources (usually upper limits). 
The latter comes through the absorption seen in TeV spectra of blazars as high energy photons
pair produce off of infrared background photons. 

As is clear from Figure~1, we have not come close to achieving any reasonable estimate of the EBL spectrum over 
three decades in wavelength  from 0.1 $\mu$m to 100 $\mu$m which is in stark contrast to the progress achieved
at X-ray energies between $\sim$2$-$10\,keV.   The few attempts at absolute measurements  involve DIRBE on COBE in
several bandpasses between 1.25 $\mu$m and 240 $\mu$m [7], IRTS, a small JAXA mission, 
between 1 and 4 $\mu$m [14],   and FIRAS above 200 $\mu$m [15]. 
Because DIRBE’s confusion limit was ~5th magnitude at 2.2 $\mu$m, all recent EBL measurements 
using DIRBE require subtraction of stellar light using ancillary measurements, such as 2MASS [5].
While the HST has been used for optical [16] and UV [17] EBL measurements, the instrument was not designed for absolute
photometry and required a careful subtraction of instrument  emission and baselines (e.g. dark current) and those measurements are subject to large
uncertainties. Between $\sim$ 10 $\mu$m to 100 $\mu$m, there is a dearth of accurate measurements and the EBL
is loosely constrained from GeV/TeV source spectra, ISO/Spitzer, and with large upper limits from DIRBE.

Among the indirect methods, the large difference in  EBL estimates from TeV spectra is due to
uncertainty in the intrinsic measurements is subject to the assumed intrinsic shape of the high-energy emission spectrum, which is known 
to be time-dependent and vary from source to source. 
Furthermore, recent Fermi results of high energy blazars seem to suggest that the adopted gamma-ray spectrum of these blazars
is much softer than previously thought. This translates directly to an error on the inferred EBL constraints. 
One attempt at constraining the near-IR EBL indirectly with TeV blazars [18] (line in Fig.~1 between 1 and 4 $\mu$m)
seems to suggest that we have essentially resolved almost all of the EBL with known galaxy counts.
A second attempt [19] (shaded region in Fig.~1),  consistent with large DIRBE EBL estimates, imply that we are far away from
resolving the background, even with JWST due to source confusion.

In retrospect,  if the EBL spectrum at IR wavelengths was exactly known, then instead of indirect limits on the IR EBL with TeV spectra 
with an assumed intrinsic spectrum,  TeV sources detected in experiments such as HESS, VERITAS, and MAGIC can be used to pursue the primary goal of
understand the astrophysics of high energy emission and AGNs.  This requires that the EBL spectrum between 4 and 10 $\mu$m be determined directly
as the  attenuation of high energy photons is dependent on the shape of the spectrum in this small wavelength range where no measurements currently 
exist [20]. 

At far-IR wavelengths, cosmological surveys with Herschel are expected to resolve about 10\% of the FIRAS EBL between 250 $\mu$m and 
550 $\mu$m, though the FIRAS EBL measurement itself is largely uncertain and must be improved in the future.  
In fact, due to the negative k-correction at sub-millimeter wavelengths, the EBL at $\sim$850$\mu$m provides the strongest constraint
on the fraction of star-formation at $z>3$ which is obscured by dust.
Similar to the situation with JWST, source counts from deep surveys expected with ALMA at wavelengths greater than 300 $\mu$m  
will only be useful in understanding the importance of discrete source populations if they are accompanied by EBL measurements.

{\bf Is there a diffuse component to the EBL?}
As discussed, the integrated galaxy counts fall short of the absolute EBL estimates 
measured by DIRBE at 2.2 and 3.6 $\mu$m where there is a minimum in the zodiacal light between scattered and thermal components [21].
While it is possible that deep extragalactic survey counts miss flux from  extended components, especially in ground-based infrared surveys,
and the large difference around 1 $\mu$m based on IRTS is far more likely to be associated with residual zodiacal light [22], 
the current state of measurements do not exclude the possibility of a  faint, diffuse near-infrared background of cosmological origin.  
The issue of cosmic variance is also key since the best measures of the 1.1 and 1.6$\mu$m integrated galaxy light come from the 
NICMOS UDF [23], which spans an angular scale about $\sim$ 6 Mpc, 
smaller than the clustering scale of typical $\sim$10$^{12-13}$\,M$_{\odot}$ dark matter halos at $z >3$. 
A mission outside the zodiacal cloud with a factor of 100 lower zodi 
foreground can resolve the current discrepancies and search for diffuse background components at very low levels.
While Spitzer and HST have revealed the $z < 4$ universe in detail and JWST will detect higher redshift sources, 
future absolute EBL measurements between 0.5 $\mu$m and 2.5 $\mu$m in several narrow wavelength bandpasses with an uncertainty less 
than $0.5$ nW m$^{-2}$ sr$^{-1}$ (a factor of $\sim$ 30 to 50 improvement over DIRBE) will be necessary for this purpose.

At far-infrared wavelengths, a component of diffuse EBL arising from dust in the IGM would be a significant problem for
dark energy measurements using the next generation of supernova surveys. This grey-dust, which it has been argued would need to 
have extinction properties different from Galactic dust [24], would obscure the light from supernova and mimic
the cosmological constant providing a floor to the uncertainty in measuring the equation of state $w(z)$.

{\bf When did first stars appear?} Current theoretical models 
of reionization, when matched to Lyman-break galaxy luminosity functions out to redshifts of 7, 
suggest that first-light sources prior to full reionization contributed $\sim 1$ nW m$^{-2}$ sr$^{-1}$ to the EBL  intensity at
near-IR wavelengths. Their contribution to the lower wavelength optical 
background,  however, is negligible as the light is redshifted beyond the optical. {\it This spectral signature in the optical to near-IR EBL
should be a primary target for a dedicated mission to measure EBL at distances around 5 AU.} 

If absolute measurements of the EBL spectrum can be achieved with
errors better than 0.1 nW m$^{-2}$ sr$^{-1}$ between 0.8 $\mu$m and 2 $\mu$m, we can in fact use the amplitude and shape
of the spectral signature with JWST counts to determine when sources first turned on. The WMAP optical depth to electron scattering 
that is routinely quoted as when the reionization ended  does not provide this information uniquely
and the only other avenue to extract such information is the use of 21-cm background.

While absolute EBL spectrum is currently uncertain, fluctuations in the EBL can be used to study faint, unresolved sources.
Fluctuation measurements with HST and Spitzer have constrained, for the first time,
the surface density of faint,   unresolved sources from intermediate to high redshifts using any 
observational windows available [25].

The limited range of angular scales provided by Spitzer and HST and planned studies during the Spitzer Warm Mission in a handful of deep, but narrow, fields
complicate detailed inferences from clustering of unresolved fluctuations. 
Measurements that span out to several degree angular scales are necessary to separate cosmological sources of interest from those in the foreground, 
including tens of degree-scale correlations expected from both zodiacal and Galactic dust. 
The limitation due to small area is  worse for current fluctuation measurements over the 1 $\mu$m to 2 $\mu$m wavelength range of 
interest where first-light galaxies peak, but some limited improvements are expected with WFC3 on HST.

With the expected background from reionization estimated to be around 0.6 to 1.0 nW m$^{-2}$ sr$^{-1}$ between 1.0 $\mu$m to 1.6 $\mu$m, 
to detect few percent rms fluctuations expected at
sub-degree angular scales, planned surveys must allow anisotropy measurements to reach a rms fluctuation level at or better than
0.05 nW m$^{-2}$ sr$^{-1}$ at 30 arcminute scales or the angular power spectrum $\ell^2C_\ell/2\pi$ at $\ell \sim 100$ to 1000 down to 
10$^{-3}$ nW$^2$ m$^{-4}$ sr$^{-2}$.

Interpretation of Spitzer and WFC3 fluctuation measurements with deep surveys will be enhanced if these measurements are accompanied by
multi-wavelength studies of  clustering of resolved faint sources.
New opportunities could be available in the next decade with facilities that will effectively allow wide, deep imaging at near-IR wavelengths. 
For those surveys to accurately determine a  diffuse component and to complement deep galaxy counts 
from JWST, future space-based IR imaging must have imaging capabilities 
that span at least a square degree between 0.8 $\mu$m to 2.0 $\mu$m. The studies must be
performed in deep imaging data with sources resolved down to a few hundred nJy level.
A multi-wavelength strategy combining luminosity functions, clustering of resolved sources,
and cross-correlation measurements against other cosmological tracers (e.g., 21-cm) can be used to  extract astrophysical details  of the 
unresolved population, including the redshift distribution and to connect source distribution to that of dark matter through 
approaches such as the halo model.

Unlike the case with optical and near-IR data from space-based imagers, only a handful of attempts on fluctuation measurements exist at
far-IR wavelengths [26]. This situation, however, is expected to soon change with wide-field imaging with instruments aboard Herschel. 
In fact, Herschel fluctuation measurements are necessary to study properties of sources that produce the bulk of the background light at sub-millimeter wavelengths.
Given the recombination lines from cosmological H and He at redshifts prior to dark ages, 
with resolved sources removed, a  detailed fluctuation study of the sub-millimeter background with a followup mission to Herschel
could potentially be used to extract the detailed history of cosmological recombination 
(and any departures from standard physics) beyond the information provided by the CMB anisotropy spectrum.

{\bf What is the total radiative content of the Universe?}
We return to our central question. The reason that we do not yet know the answer to this question
is essentially an issue of foregrounds in EBL measurements, namely stars,  
interstellar dust, interplanetary dust emission, at mid-infrared and far-infrared wavelengths, and sunlight scattered by 
interplanetary dust (‘zodiacal light’), at optical and near-infrared wavelengths.  
Many DIRBE-based and IRTS-based results use a model of interplanetary dust scattering and emission [27], 
while others use an alternate model based on the principle that the zodiacal residual be zero at 25 $\mu$m at high ecliptic latitudes [28].  
The HST results on the optical background subtract zodiacal emission based on the 
observed strength of reflected Fraunhofer lines from a ground-based measurement [16].  
Unfortunately, zodiacal dust models are not unique and represent a leading source of systematic error. In the shortest 
DIRBE band, the difference between the two leading zodi models is somewhere between 50\% to 100\% of the EBL estimates 
($\sim 20$ to 40 nW m$^{-2}$ sr$^{-1}$)  at near-IR wavelengths. {\it Simply repeating an absolute photometry experiment like DIRBE or IRTS 
will not improve the current EBL spectrum at $\lambda <5$ $\mu$m, 
unless supplemented with on-board absolute spectroscopy to monitor scattered Fraunhofer lines.} 

Improvements in EBL measurements must come with a parallel improvement in our understanding of the zodiacal cloud, including 
the structure of the zodiacal cloud and the distribution of zodiacal dust particles over their orbital
elements. The structure and the density profile depend on sources of dust particles and sizes of the particles and it is still not established
whether asteroidal or cometary dust dominate in the zodiacal cloud [29]. 
{\it In situ} measurements with Pioneer 10 indicate that dust density is dropping more
rapidly than the $1/r$ radial behavior expected from the Poynting-Robertson force [30].
The zodiacal cloud itself contains unique science involving dynamics within the Solar system.
The out-of-zodi EBL measurements (proposed below) and studies of the zodi velocity profiles, together
with models of migration of dust, will help us to understand better the
distribution of dust particles beyond Jupiter distance over their orbital elements
and to estimate fractions and typical sizes of cometary and trans-Neptunian particles in the zone of the giant planets.

{\bf Priorities for 2010-2020}
The absolute sky brightness is so strongly dominated by zodiacal light from visible to mid-infrared wavelengths that it is not possible to measure the EBL 
with confidence at distances of 1 AU with a 0.1\% or better removal of zodi needed to detect the reionization spectral feature around 1 $\mu$m (Figure~1 right panel).
The only way to measure the EBL with any useful accuracy between 5 and 50 $\mu$m is to conduct measurements outside the zodiacal cloud.
There are two possibilities: traveling beyond Jupiter, or above the ecliptic plane.
At distance of Jupiter, existing {\it in-situ} measurements with Pioneer 10 indicate a decrease in
zodiacal light of two orders of magnitude, relative to the brightness at 1 AU. In addition to EBL
measurements, on the way to the outer Solar System, measurements can be made to establish
the dust density radial and azimuthal profiles. A measurement of the EBL and the interplanetary dust distribution
can be conceived either as a modest mission dedicated to this purpose or as a camera that piggy-backs on a Planetary mission
to the outer Solar System.

As first steps towards this priority sounding rocket experiments can be pursued for absolute photometry [31].
Methods of constraining the zodiacal light simultaneously, e.g. using Fraunhofer  lines as a spectral signature, have been developed and used successfully in the past. 
The time variation of zodi in three to six month intervals along the same lines of sight can be pursued to extract additional information.  

\begin{multicols}{2}
\begin{scriptsize}
\noindent
[1] Hauser, M.~G. \& Dwek, E. 2001, ARA\&A, 39, 249; Kashlinsky, A. 2005, Phys. Rep.  409, 361

\noindent
[2] Lagache, G., Puget, J.-L. \& Dole, H. 2005, ARA\&A, 43, 727

\noindent
[3] Pei, Y.~C., Fall, S.~M. \& Hauser, M.~G. 1999, ApJ, 522, 604

\noindent
[4] Levenson, L. \& Wright, E., 2008, ApJ, 683, 585

\noindent
[5] e.g., Wright, E.~L. 2001, ApJ, 553, 538; 

\noindent
[6] Dole, H. {\it et al.} 2006, A\&A, 451, 417

\noindent
[7] Hauser, M.~G. {\it et al.} 1998, ApJ, 508, 25

\noindent
[8] Ferguson, H. {\it et al.}  2002, ApJ, 569, 65

\noindent
[9] e.g., Santos, M.~R., Bromm, V. \& Kamionkowski, M. 2002, MNRAS, 336, 1082; Kashlinsky, A. {\it et al.} 2004, ApJ, 608, 1; 
Cooray, A. {\it et al.} 2004, ApJ, 606, 611; Fernandez, E. \& Komatsu, E. 2006, ApJ, 646, 703  

\noindent
[10] e.g., Iliev, I. {\it et al.} 2008, MNRAS, 391, 63

\noindent
[11] Cooray, A. \& Yoshida, N. 2004, MNRAS, 351, L71

\noindent
[12] Feng, J. {\it et al.} arXiv.org:0704.1658

\noindent
[13] Wong, W.~Y., {\it et al.} 2006, MNRAS, 367, 1666

\noindent
[14] Matsumoto, T. {\it et al.} 2005, ApJ, 626, 31 

\noindent
[15] Mather, J.~C. {\it et al.} 1994, ApJ, 420, 439

\noindent
[16] Bernstein, R.~A. 2007, ApJ, 666, 663

\noindent
[17] Brown et al. AJ, 120. 1153

\noindent
[18] Aharonian, F. {\it et al.}, 2006, Nature, 440, 1018 

\noindent
[19] Schroedter, M. 2005, ApJ, 628, 617

\noindent
[20] Mapelli, M. {\it et al.} 2006, New Astro., 11, 420  

\noindent
[21] Madau, P. \& Pozzetti, L. 2000, MNRAS, 312, L9

\noindent
[22] Dwek, E., Arendt, R., Krennrich, F. 2005, ApJ, 635, 784; 

\noindent
[23] Thompson, R. {\it et al.} 2007, ApJ, 657, 669

\noindent
[24] Aguirre, A. {\it et al.} 1999, ApJ, 525, 583

\noindent
[25] Kashlinsky, A. {\it et al.} 2005 Nat, 438, 45; 2007, ApJ, 666, L1; Cooray, A. {\it et al.} 2007, ApJ, 659, L91; Thompson, R. {\it et al.} 2007, 666, 658; Chary, R. {\it et al.} 2008, ApJ, 681, 53 

\noindent
[26] Lagache, G. {\it et al.} 2007, ApJ, 665, L89; Grossan, B. \& Smoot, G~F. 2007, A\&A, 474, 731

\noindent
[27] Kelsall, T. {\it et al.} 1998, ApJ, 508, 44

\noindent
[28] Gorjian, V., Wright, E.~L. \& Chary, R.~R. 2000, ApJ, 536, 550 

\noindent
[29] Ipatov, S. {\it et al.}, 2008, Icarus, 194, 769

\noindent
[30] Hanner, M. {\it et al.} 1974, JGR, 79, 3671 

\noindent
[31] Bock, J. {\it et al.}, 2006, NewA Rev. 50, 215

\end{scriptsize}
\end{multicols}

\end{document}